\documentclass[12pt]{article}
\usepackage{a4wide}
\usepackage{epsfig}

\newcommand{\GeV}{\mathop{\rm\,GeV}\nolimits}
\newcommand{\MeV}{\mathop{\rm\,MeV}\nolimits}
\newcommand{\Disc}{\mathop{\rm Disc}\nolimits}
\newcommand{\msbar}{\overline{\mbox{\rm MS}}}
\newcommand{\eps}{\varepsilon}
\newcommand{\pfrac}[2]{\left(\frac{#1}{#2}\right)}
\newcommand{\slx}{x\kern-5.5pt\raise1pt\hbox{$\scriptstyle/$}\kern1.5pt}
\begin{document}
\thispagestyle{empty}
\begin{flushright}
MZ-TH/11-17\\
SI-HEP-2011-06\\
July 2011
\end{flushright}

\begin{center}
{\Large\bf Calculating loops without loop calculations:\\[0.3truecm]
           NLO computation of pentaquark correlators}\\[0.5truecm]

{\large S.~Groote,$^{1,2}$ J.G.~K\"orner$^1$ and
  A.A.~Pivovarov$^{1,3,4}$}\\[.4cm]
$^1$ Institut f\"ur Physik, Johannes-Gutenberg-Universit\"at,\\[.1truecm]
  Staudinger Weg 7, D-55099 Mainz, Germany\\[.3truecm]
$^2$ F\"u\"usika Instituut, Tartu \"Ulikool,
  Riia~142, EE--51014 Tartu, Estonia\\[.3truecm]
$^3$ Institute for Nuclear Research of the\\[.1truecm]
  Russian Academy of Sciences, Moscow 117312, Russia\\[.3truecm]
$^4$ Theoretische Physik 1, Universit\"at Siegen,\\
Walter-Flex-Strasse 3, D-57068 Siegen, Germany
\end{center}

\begin{abstract}
We compute next-to-leading order (NLO) perturbative QCD corrections to the
correlators of interpolating pentaquark currents and their absorptive parts.
We employ modular techniques in configuration space which saves us from the
onus of having to do loop calculations. The modular technique is explained in
some detail. We present explicit NLO results for several interpolating
pentaquark currents that have been written down in the literature. Our modular
approach is easily adapted to the case of NLO corrections to multiquark
correlators with an arbitrary number of quarks/antiquarks.
\end{abstract}

\newpage
\section{Introduction}
The discovery of exotic quark states and bound states of gluons would be 
another manifestation of QCD, allowing for a quantitative check of its features
and finding numerical values of some important QCD parameters. While glueballs
are certainly the most searched for states in QCD~\cite{Brodsky:2003hv}, there
is also much interest in exotic states in strong interactions, i.e.\ states
built from quarks within QCD which differ from the simplest valence quark
content of mesons or baryons (see e.g.\ Refs.~\cite{Jaffe,Matveev}). The
theoretical investigation of multiquark states ($Q^n\bar Q^m$, $n+m>3$) and
the experimental search for them may provide important information on the
properties of the interaction of quarks and gluons at large distances. Until
recently major efforts have been directed to the study of the dibaryon
spectrum ($n=6$, $m=0$), both theoretically and experimentally~\cite{Yokosawa}.
This particular six-fold state is rather peculiar as it is close to the
deuteron, building bridges to applications of QCD to medium-energy nuclear
physics~\cite{Larin:1985gb,nuclear}. In Ref.~\cite{Jaffe}, Jaffe predicted
that there might exist a stable six-quark $S$-wave state -- a dihyperyon $H$
-- which is a singlet with respect to both colour and flavour $SU(2)$ (with
strangeness $-2$) with the quantum numbers $J^P=0^+$ and a mass around
$2150\MeV$. The quantum numbers of the $H$ state are identical to the quantum
numbers of the $(\Lambda\Lambda)$ pair of two $\Lambda(1115)$ hyperons, and
its mass is smaller than the sum of the masses of the two $\Lambda$ hyperons.
The $H$ state is therefore stable with respect to strong interactions and can
decay only through weak interactions. To the best of our knowledge, the famous
dibaryon state $H$ is the first state to attract attention in the modern
context of QCD. Thereafter there were efforts to identify some mesons in QCD
(scalar mesons as a $KK$ molecule) with a four quark state in order to explain
their properties and, in particular, their production and decay
patterns~\cite{Braun:1988kv}. In the intervening years the interest in exotic
states has mainly shifted to tetraquarks and pentaquarks.

The study of bound states in QCD is a difficult problem. After almost 40 years
of research it is clear that the most promising approach is very likely given 
by lattice QCD, in particular, since the computer power and computer
algorithms have advanced much since the first introduction of lattice QCD in
the early seventies of the last century (results are given for instance in
Ref.~\cite{Wetzorke:2002mx}). Besides lattice QCD, model dependent approaches
have been used in Refs.~\cite{models,Diakonov:1997mm}, for example, in the
framework of the MIT quark-bag model~\cite{Chodos}. It is important to test
these model predictions solely on the basis of fundamental principles of QCD.
Such a test can be made by means of the method of QCD sum rules, using either
the technique of finite-energy sum rules~\cite{Chetyrkin} or that of Borel sum
rules~\cite{Shifman}.

The operator product expansion and QCD sum rules serve as a solid testing 
ground for many calculations in the theory of hadrons. The method of QCD sum
rules is based on the fundamental field theoretic principles of QCD, and has
proved its effectiveness in calculations of the masses of
mesons~\cite{Shifman,mesonsums,Kras1} and baryons~\cite{Ioffe,Chung:1981cc,%
Espriu:1983hu}. However, the reliability of perturbative calculations
requires a thorough check, in particular, in the uncharted territory of exotic
multiquark states where the collected experimental material is rather small.
It is therefore worthwhile to compute some examples in order to get a feeling
for the structure of the perturbative series. Work in this direction is under
way.

Glueballs have been previously analyzed in the context of QCD sum rules. The  
perturbative QCD corrections to the sum rules were found to be very 
large~\cite{Kat1}. Exotic mesonic states have been analyzed in
Refs.~\cite{exmesons}. QCD sum rules for ordinary three quark baryon states
have been widely studied. In particular, the correlators of baryonic currents
with finite mass heavy quarks have been calculated at next-to-leading order of
perturbative QCD, allowing for further improvements in the precision of QCD
sum rule predictions~\cite{nlosums}. It is known that next-to-leading order
(NLO) perturbative corrections to baryon sum rules are
large~\cite{Kras1,nlosums,Jamin:1987gq}. They are expected to be even larger
for multiquark states with $n>3$ quarks. Different aspects of such $n>3$
multiquark states in QCD have already been discussed some time
ago~\cite{Larin:1986yt}. One feature of $n>3$ multiquark states is that they
have a large internal weight of colour states~\cite{Matveev}.

The immediate purpose of the present investigation is to concentrate on a type
of exotic multiquark state called pentaquarks -- states with baryon quantum
numbers that contain an additional quark-antiquark pair. These states have
been discovered experimentally by different collaborations: LEPS Collaboration
(Japan)~\cite{Nakano:2003qx}, DIANA Collaboration
(Russia)~\cite{Barmin:2003vv}, CLAS Collaboration
(USA)~\cite{Stepanyan:2003qr}, and SAPHIR Collaboration
(Germany)~\cite{Barth:ja}. The results of the present investigation will open 
the possibility for a high-precision description of these experimental data on
pentaquarks. The investigation is also important for further experimental
precision studies on these and related states at DESY (HERMES
Collaboration~\cite{Airapetian:2003ri}) and CERN (NA49
Collaboration~\cite{Alt:2003vb}). There is also a proposal to launch an
experimental study of pentaquark baryons at meson
factories~\cite{Browder:2004mp}. Experimentally these collaborations are using
different apparata and techniques but theoretically the observed states should
be understood within QCD. While the first principle numerical computation 
on the lattice gave rather positive results~\cite{Sasaki:2003gi}, analytical
methods and in particularly method of QCD sum rules should definitely be
developed for a reliable identification of the new states in the hadronic
spectrum.

It appears that the experimental confirmation of these states is problematic
at the moment as some collaborations have reconsidered their results and
conclusions. However, there is no doubt that such states are possible within
QCD and the theoretical study should continue. In case of a definite positive
indication from theory the experimental searches could certainly proceed in a
much more efficient way.

In particular, a dedicated experiment has given a negative result in the direct
search of the pentaquark state~\cite{Battaglieri:2005er}. A review of the
present experimental situation can be found in
Refs.~\cite{experiment} (see also Ref.~\cite{Barnes:2005zy}). Note that more
lattice studies have become available~\cite{lattice}, some with a negative
outcome as concerns the existence of pentaquark states~\cite{Holland:2005yt}.
This makes the task of the theory even more challenging. Either one has to
show that such states do not form for some reason, or to suggest a new mass
scale of these states and to identify the appropriate decay modes for their
determination~\cite{decaymodes}. The first task is difficult in as much as one
touches on the problem of bound state formation and therefore of the
(confined) strong coupling. The latter problem ultimately requires the
calculation of perturbative corrections to the operator product expansion used
within QCD sum rules.

Sum rule calculations of pentaquarks and corresponding critical analysis'
have been presented in numerous papers~\cite{Zhu:2003ba,pentasums,%
Matheus:2004gx}. While the accuracy of the QCD sum rule method is about
$\sim 20\%$ at present, the results obtained agree with experimental claims
and model predictions~\cite{Diakonov:1997mm}. However, within the QCD sum rule
method it is not possible to predict whether the mass of the lowest pentaquark
state lies above or below the $Kp$ threshold (i.e., whether it is stable).

The aim of the present paper is to create a framework for an accurate sum rule
analysis of the properties of pentaquark states. The study of pentaquark
states within the QCD sum rule method requires a precise knowledge of the
absorptive parts of the correlators of the pentaquark interpolating currents.
In this paper we present perturbative next-to-leading order calculations of
the relevant correlators in QCD.

\section{NLO corrections to the correlation function}
According to the QCD sum rule approach to hadron properties, the principal 
quantity to be analyzed is the correlation function,
\begin{equation}\label{corrfun}
\Pi(q)=i\int d^4x e^{iqx}\langle 0|Tj(x)\bar j(0)|0\rangle,
\end{equation}
where $j(x)$ is a local current operator with the quantum numbers of the
hadron state, termed the interpolating current of the hadron state. The 
construction of the conjugate operator $\bar j(x)$ depends on whether the
hadron is a fermion or a boson. For fermionic states such as the ordinary
baryon states or the pentaquark states dealt with in this paper, one has
$\bar j(x)=j^\dagger(x)\gamma^0$. For bosonic states (mesons, tetraquarks,
\dots) the conjugate operator is just the adjoint operator,
$\bar j(x)=j^\dagger(x)$. The result of the sum rule analysis depends strongly
on the choice of the interpolating current as has been shown already in the
case of the dibaryon state~\cite{Larin:1986yt}.

In the case of a given pentaquark state the pentaquark current $j(x)$ is a 
local scalar current with the quantum numbers of that pentaquark baryon. For
instance, take the ground state pentaquark state $\Theta^+$. The current is
constructed from five quark fields, such that its projection onto the real
pentaquark baryon state $|\Theta^+(p)\rangle$  (within the assumption that
this state exists) is nonzero:
\begin{equation}
\langle 0|j(0)|\Theta^+(p)\rangle =\lambda_{\Theta^+}, 
\quad p^2=m_{\Theta^+}^2.
\end{equation}
Since such a current $j(x)$ is not unique, the question of its optimal choice
arises immediately (see Appendix~B for a discussion of this issue). We recall
that the problem of choosing the current already arose in the case of
baryons~\cite{Ioffe,Chung:1981cc} where the currents are constructed
from three quark fields. When the current is constructed from five quark
fields as in our case, this problem is much more complicated, since the number
of independent currents with the given quantum numbers is much larger (see also
Ref.~\cite{Matheus:2004gx}). The treatment of the current $j(x)$ in the most
general form, i.e.\ in the form of a linear combination of all the independent
local operators with the quantum numbers of $\theta$, is a very cumbersome
problem. Therefore we confine ourselves to the choice of a few of the simplest
currents with the required quantum numbers and analyze the dependence of our
results on the properties of these currents. As in the case of mesons and
baryons, we shall construct the current $j(x)$ from quark fields without
derivatives.  

In accordance with the method of QCD sum rules, we shall calculate the
correlation function~(\ref{corrfun}) by means of Wilson's operator expansion,
assuming that the vacuum expectation values of the local operators (the
so-called condensates) are nonzero. The calculations must be performed in the
Euclidean region $-q^2\ge 1\GeV^2$. In this region, the effective strong
interaction constant $\alpha_s$ is not very small and one has to calculate the
coefficient functions of the operator expansion at least at NLO in
perturbation theory. Therefore, the correlator function $\Pi$ should be
calculated at NLO in $\alpha_s$. The fact that this may be necessary for
calculations of physical quantities in the framework of the sum rule method is
confirmed by previous applications of the sum rule method, in particular, by
the calculation of baryon masses. 
 
In this paper we explicitly discuss the computational techniques for the unity
operator of the operator product expansion. The condensate contributions to
the correlation function which have to be incorporated for a consistent NLO
analysis are not discussed in this paper but can be calculated along the lines
presented here. For instance, the incorporation of the quark condensate
requires only minor modifications of the present methods. Quark condensate
contributions to baryonic sum rules have been e.g.\ considered in 
Ref.~\cite{Groote:2008hz}.

\begin{figure}[t]\begin{center}
\includegraphics[angle=0,width=0.25\textwidth]{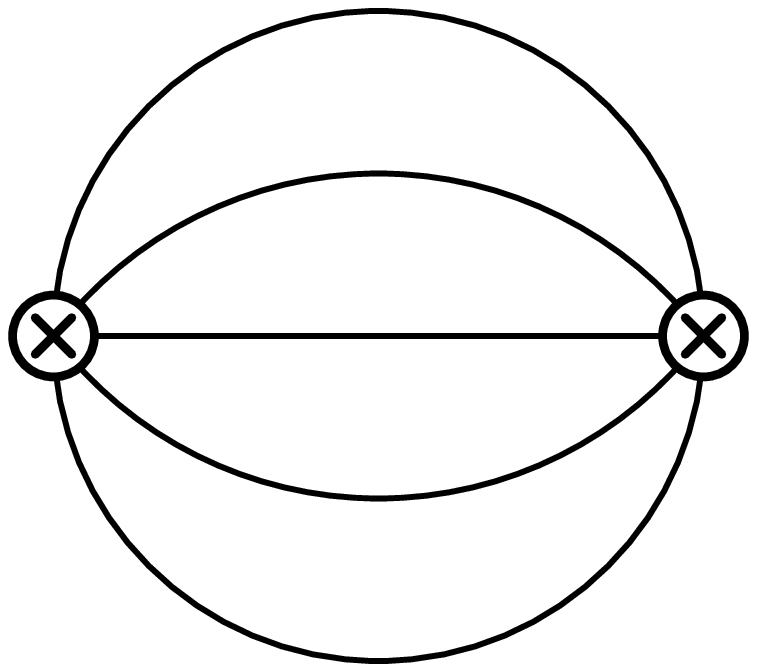}
\includegraphics[angle=0,width=0.25\textwidth]{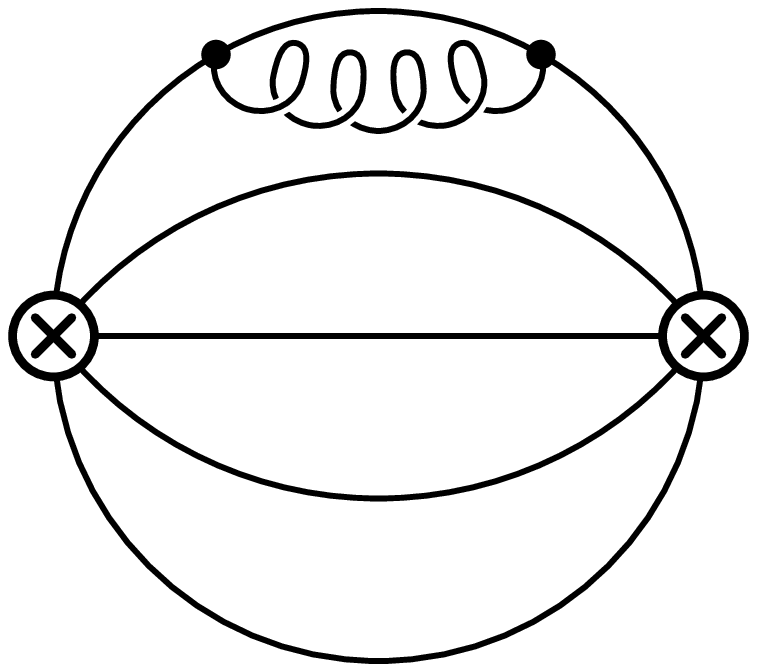}
\lower4pt\hbox{\includegraphics[angle=0,width=0.25\textwidth]{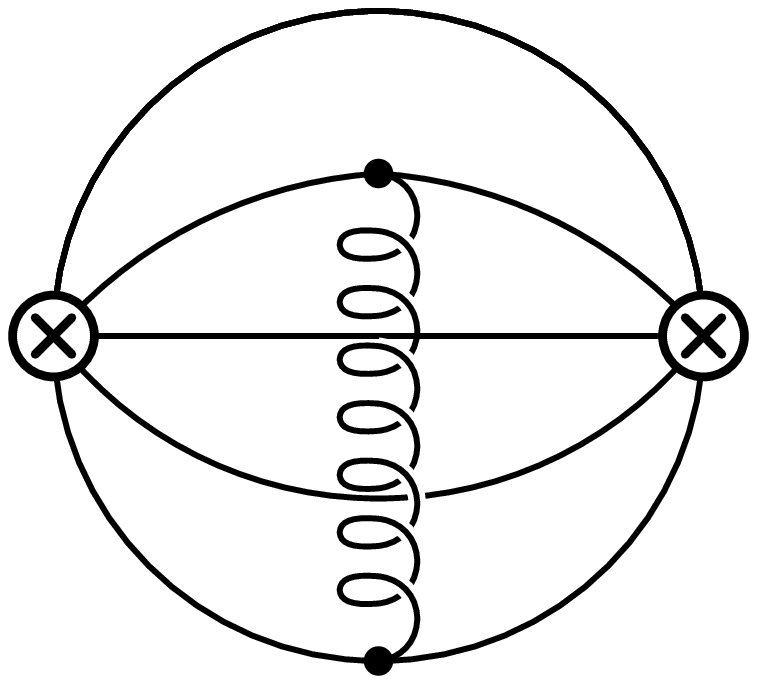}}
\vspace{7pt}
\centerline{(a)\kern104pt(b)\kern104pt(c)}
\caption{\label{pentcor}LO contribution (a) and examples for the NLO propagator
 (b) and dipropagator corrections (c).}
\end{center}\end{figure}

The LO calculation falls into the category of the sunset diagrams (cf.
Fig.~\ref{pentcor}a). Sunset diagrams are directly calculable in configuration
space~\cite{sunsets}. These types of diagrams also appear in the effective low
energy gluon correlator for light quarks~\cite{gluoncor}. The corrections are
of two types. The propagator-type corrections depicted in Fig.~\ref{pentcor}b
are straightforward and can be easily added with no effort at all. The second
type of corrections depicted in Fig.~\ref{pentcor}c correspond to the
irreducible diagrams of the fish type. They are rather well known in the
massless limit (the more complicated massive case was analyzed in
Ref.~\cite{Narison:1994zt}). In order to deal with the diversity of
interpolating currents that have been proposed in the literature for the
pentaquark states, we have developed a modular calculation method in
configuration space. The modular method reduces the perturbative calculations
of the present paper to pure algebraic calculations~\cite{Groote:2006sy}.

\section{Presentation of the modular method}
As already mentioned, the two required main modules of our method are the 
propagator correction $S_1(x)$ and the dipropagator correction $S_2(x)$ which
read ($\slx=\gamma^\mu x_\mu$)
 \begin{eqnarray}
\lefteqn{S_1(x)|_{\rm NLO}\ =\ S_1(x)|_{\rm LO}
\left\{1-C_F\frac{\alpha_s}{4\pi\eps}\left(\mu_x^2x^2\right)^\eps
  \right\}\ =\ S_0(x^2)\left\{1-C_F\frac{\alpha_s}{4\pi\eps}
  \left(\mu_x^2x^2\right)^\eps\right\},}\nonumber\\
  \label{prop}\\
\lefteqn{S_2(x)|_{\rm NLO}\ =\ S_0(x^2)^2
  \Bigg\{\slx\otimes\slx
  -t^a\otimes t^a\frac{\alpha_s}{4\pi}(\mu_x^2x^2)^\eps\times\strut}
  \nonumber\\&&\strut\Bigg(\gamma^\mu\otimes\gamma^\nu
  \Bigg[\left(\frac1\eps+\frac{11}2\right)x_\mu x_\nu
  +\left(\frac1\eps+\frac12\right)x^2g_{\mu\nu}\Bigg]
  +\left(\frac1{2\eps}+\frac14\right)\Gamma_3^{\alpha\beta\mu}
  \otimes{\Gamma_{3\ \alpha\beta}}^\nu x_\mu x_\nu\Bigg)\Bigg\}\qquad
  \label{diprop}
\end{eqnarray}
where in the Euclidean domain one has
\begin{equation}\label{S0def}
S_0(x^2)=\frac{-\Gamma(2-\eps)}{2\pi^{2-\eps}(x^2)^{2-\eps}}.
\end{equation}
The renormalization scale $\mu_x$ is appropriate for calculations in
configuration space if one wants to avoid the appearance of $\ln(4\pi)$ and 
$\gamma_E$ terms. The scale $\mu_x$ is related to the scale $\bar\mu$ of the 
$\msbar$ scheme by
\begin{equation}
\mu_x=\bar\mu e^{\gamma_E}/2.
\end{equation}
The direct product signs ``$\otimes$'' in the dipropagator correction $S_2(x)$
serve to distinguish between the two fermion lines involved in the gluon
exchange. Finally,
\begin{equation}
\Gamma_3^{\mu\alpha\nu}=\gamma^{[\mu}\gamma^\alpha\gamma^{\nu]}=\frac12
  (\gamma^\mu\gamma^\alpha\gamma^\nu-\gamma^\nu\gamma^\alpha\gamma^\mu)
\end{equation}
is the totally antisymmetric product of three gamma matrices. Equations
(\ref{prop}) and~(\ref{diprop}) allow one to calculate the corrections to
$n$-quark/antiquark current correlators of any composition without having to
calculate any integrals. In Ref.~\cite{Groote:2006sy} we have presented
results for a model current with five different flavours. In this paper we
deal with several interpolating currents suggested in the literature including
the equal flavour case. Because of flavours appearing twice or more times in
the interpolating current, the Wick contraction will result in a main
contribution and different ``crossover'' contributions.

Before giving our results for the various interpolating currents, we have to
deal with renormalization. Corresponding to the propagator and dipropagator
corrections, the correlator function is renormalized by the wave function
renormalization factor and the vertex renormalization factor, respectively.
Concerning the vertex renormalization factor one has to account for mixing
effects. Mixing can occur when gluons are exchanged between quark lines in the
pentaquark correlation function. Mixing is taken into account through the
subtraction of corresponding vertex divergences generated by an operator that
can admix to the initial current. The general formula reads
\begin{equation}
(\psi_i\otimes\psi_j)_R=(\psi_i\otimes\psi_j)
  -\frac{\alpha_s}{4\pi\eps}(1_{ii'}\otimes 1_{jj'}
  +\frac14\sigma^{\alpha\beta}_{ii'}\otimes\sigma^{\alpha\beta}_{jj'})
(\psi_{i'}\otimes\psi_{j'}).
\end{equation}
Here $\sigma^{\alpha\beta}=i/2[\gamma^{\alpha},\gamma^{\beta}]$ and 
all numbers are calculated in Feynman gauge. Note that the part proportional
to $\sigma$ is gauge independent. The renormalization within our modular
approach follows from the above line of arguments and leads to counterterms
which are listed in explicit form in the following.

Once the correlator function is renormalized, we can calculate the spectral
density corresponding to the correlator. For this purpose, instead of
calculating explicitly via
\begin{equation}
\Pi(q)=i\int d^4x e^{iqx}\langle 0|Tj(x)\bar j(0)|0\rangle.
\end{equation}
in momentum space, one can use the formulas given in Appendix~A.

\section{Results for pentaquarks of the first kind}
The correlators in this section are
\begin{equation}
\langle 0|Tj(x)\bar j(0)|0\rangle=S_0(x^2)^5(x^2)^2\slx\Pi_j(x^2).
\end{equation}
We start with different interpolating currents proposed for the lowest
pentaquark state $\Theta^+$ at $1530\MeV$ with quantum numbers $J^P=1/2^+$ and
$S=1$. Reference~\cite{Zhu:2003ba} gives an overview over pentaquarks which
are built up by a diquark, a meson and a single quark. In the following these
currents will be called pentaquark currents of the first kind. The
interpolating current with isospin $I=0$ is given by
\begin{equation}
\eta_0(x)=\frac1{\sqrt2}\epsilon_{abc}\left[u_a^T(x)C\gamma_5d_b(x)\right]
  \left\{u_e(x)\bar s_e(x)i\gamma_5d_c(x)-(u\leftrightarrow d)\right\}.
\end{equation}
Because of the two parts of the interpolating current, there are two diagonal
and two mixed bare contributions,
\begin{eqnarray}
\Pi_{\eta_0B}^{11}(x^2)\!\!\!&=&\!\!\!180\left\{1+\frac{\alpha_s}\pi
  (\mu_x^2x^2)^\eps\left(\frac1\eps+\frac{13}3\right)\right\}
  -12\left\{1+\frac{\alpha_s}\pi(\mu_x^2x^2)^\eps\left(\frac3\eps+3\right)
  \right\}\nonumber\\&&\strut
  -3\left\{1+\frac{\alpha_s}\pi(\mu_x^2x^2)^\eps
  \left(-\frac1\eps+\frac{17}3\right)\right\}=\Pi_{\eta_0B}^{22}(x^2),
  \nonumber\\
\Pi_{\eta_0B}^{12}(x^2)\!\!\!&=&\!\!\!18\left\{1+\frac{\alpha_s}\pi
  (\mu_x^2x^2)^\eps\left(\frac7\eps+\frac13\right)\right\}
  +3\left\{1+\frac{\alpha_s}\pi(\mu_x^2x^2)^\eps
  \left(-\frac1\eps+\frac{17}3\right)\right\}=\Pi_{\eta_0B}^{21}(x^2).
  \nonumber\\
\end{eqnarray}
The counterterms for the current read
\begin{eqnarray}
\Delta\Pi_{\eta_0}^{11}&=&-180\frac{\alpha_s}\pi\pfrac1\eps
  +12\frac{\alpha_s}\pi\left(\frac3\eps-\frac73\right)
  -3\frac{\alpha_s}\pi\pfrac1\eps\ =\ \Delta\Pi_{\eta_0}^{22},\nonumber\\
\Delta\Pi_{\eta_0}^{12}&=&-18\frac{\alpha_s}\pi\left(\frac7\eps-\frac{14}3
  \right)+3\frac{\alpha_s}\pi\pfrac1\eps\ =\ \Delta\Pi_{\eta_0}^{21}.
\end{eqnarray}
The singularities cancel in the renormalized results which reads
\begin{eqnarray}
\Pi_{\eta_0R}^{11}(x^2)&=&180\left\{1+\frac{\alpha_s}\pi
  \left(\frac{13}3+\ln(\mu_x^2x^2)\right)\right\}
  -12\left\{1+\frac{\alpha_s}\pi\left(\frac{16}3+3\ln(\mu_x^2x^2)\right)
  \right\}\nonumber\\&&\strut
  -3\left\{1+\frac{\alpha_s}\pi(\mu_x^2x^2)^\eps
  \left(\frac{17}3-\ln(\mu_x^2x^2)\right)\right\}
  \ =\ \Pi_{\eta_0R}^{22}(x^2),\nonumber\\
\Pi_{\eta_0R}^{12}(x^2)&=&18\left\{1+\frac{\alpha_s}\pi
  \left(5+7\ln(\mu_x^2x^2)\right)\right\}
  +3\left\{1+\frac{\alpha_s}\pi\left(\frac{17}3-\ln(\mu_x^2x^2)\right)
  \right\}\ =\ \Pi_{\eta_0R}^{21}(x^2).\nonumber\\
\end{eqnarray}
In order to calculate the spectral density we have to treat the first order
correction and the counterterm separately. The reason is that these two
contributions have different $x^2$ powers. The general procedure for the
calculation of the spectral density is left to Appendix~A. The result for the
spectral density reads
\begin{equation}
\rho(s)=\frac{s^5}{604800(4\pi)^8}
  \left\{A_0+\frac{\alpha_s}\pi\left(B_1+C_1+\frac{512}{105}B_0
  +B_0\ln\pfrac{\bar\mu^2}s\right)\right\}
\end{equation}
where $A_0$ is the LO contribution, $B_0$ and $B_1$ are the singular and finite
parts of the NLO result, respectively, and $C_0\,(\,=\,-B_0)$ and $C_1$ are the
singular and finite parts of the counterterm. Collecting all contributions
one obtains
\begin{eqnarray}
\Pi_{\eta_0B}&=&372+60\frac{\alpha_s}\pi\left(\frac9\eps+25\right)
  \ =\ A_0+\frac{\alpha_s}\pi\left(\frac{B_0}\eps+B_1\right),
  \nonumber\\
\Delta\Pi_{\eta_0}&=&60\frac{\alpha_s}\pi\left(-\frac9\eps+\frac{28}{15}
  \right)\ =\ \frac{\alpha_s}\pi\left(\frac{C_0}\eps+C_1\right).
\end{eqnarray}
The spectral density finally reads
\begin{eqnarray}
\rho_{\eta_0}(s)&=&\frac{s^5}{604800(4\pi)^8}\left\{372
  +60\frac{\alpha_s}\pi\left(9\ln\pfrac{\bar\mu^2}s+\frac{7429}{105}\right)
  \right\}\ =\nonumber\\
  &=&\frac{31s^5}{50400(4\pi)^8}\left\{1+\frac{\alpha_s}\pi
  \left(\frac{45}{31}\ln\pfrac{\bar\mu^2}s+\frac{7429}{651}\right)\right\}.
\end{eqnarray}
The perturbative correction can be seen to be rather large, cf.\
$7429/651(\alpha_s/\pi)$. For the remaining currents proposed in
Ref.~\cite{Zhu:2003ba}, one has
\begin{eqnarray}
\eta_1(x)&=&\frac1{\sqrt2}\epsilon_{abc}\left[u_a^T(x)C\gamma_5d_b(x)\right]
  \left\{u_e(x)\bar s_e(x)i\gamma_5d_c(x)+(u\leftrightarrow d)\right\},\\
\eta'_1(x)&=&\frac1{\sqrt2}\epsilon_{abc}\left[u_a^T(x)C\gamma^\mu d_b(x)
  \right]\left\{\gamma_\mu\gamma_5u_e(x)\bar s_e(x)i\gamma_5d_c(x)
  -(u\leftrightarrow d)\right\},\\
\eta_2(x)&=&\frac1{\sqrt2}\epsilon_{abc}\left\{\left[u_a^T(x)C\gamma^\mu u_b(x)
  \right]\gamma_\mu\gamma_5d_e(x)\bar s_e(x)i\gamma_5d_c(x)
  +(u\leftrightarrow d)\right\},\\[3pt]
\eta'_2(x)&=&\epsilon_{abc}\left[u_a^T(x)C\gamma^\mu u_b(x)\right]
  \gamma_\mu\gamma_5u_e(x)\bar s_e(x)i\gamma_5u_c(x).
\end{eqnarray}
Again we only give results for the spectral densities. They read
\begin{eqnarray}
\rho_{\eta_1}(s)&=&\frac{s^5}{2100(4\pi)^8}\left\{1+\frac{\alpha_s}\pi
  \left(\frac16\ln\pfrac{\bar\mu^2}s+\frac{517}{105}\right)\right\},\\
\rho_{\eta'_1}(s)&=&\frac{17s^5}{6300(4\pi)^8}\left\{1+\frac{\alpha_s}\pi
  \left(\frac{2255}{408}\right)\right\},\\
\rho_{\eta_2}(s)&=&\frac{s^5}{525(4\pi)^8}\left\{1+\frac{\alpha_s}\pi
  \left(-\frac76\ln\pfrac{\bar\mu^2}s+\frac{377}{360}\right)\right\},\\
\rho_{\eta'_2}(s)&=&\frac{s^5}{525(4\pi)^8}\left\{1+\frac{\alpha_s}\pi
  \left(-\frac{11}6\ln\pfrac{\bar\mu^2}s-\frac{15877}{10080}\right)\right\}.
\end{eqnarray}
The perturbative corrections can become as large as
$2255/408(\alpha_s/\pi)$.

\section{Pentaquarks of the second kind}
Pentaquarks of the second kind consist of two diquarks and one antiquark.
Three possible choices are given in Ref.~\cite{Sasaki:2003gi}, and, with some
small deviations, in Refs.~\cite{Oka:2004kb,Gubler:2009ay}. They read
\begin{eqnarray}
\theta_1^+(x)&=&\epsilon_{abc}\epsilon_{aef}\epsilon_{bgh}
  \left[u_e^T(x)Cd_f(x)\right]\left[u_g^T(x)C\gamma_5d_h(x)\right]
  C\bar s_c^T(x),\\[7pt]
\theta_2^{+,\mu}(x)&=&\epsilon_{abc}\epsilon_{aef}\epsilon_{bgh}
  \left[u_e^T(x)C\gamma_5d_f(x)\right]\left[u_g^T(x)C\gamma^\mu\gamma_5
  d_h(x)\right]C\bar s_c^T(x),\\[7pt]
\theta_3^{+,\mu}(x)&=&\epsilon_{abc}\epsilon_{aef}\epsilon_{bgh}
  \left[u_e^T(x)Cd_f(x)\right]\left[u_g^T(x)C\gamma^\mu\gamma_5d_h(x)\right]
  \gamma_5C\bar s_c^T(x).
\end{eqnarray}
First we consider the case when the Lorentz index $\mu$ in the correlator is
contracted. One then has
\begin{eqnarray}
\rho_{\theta_1}(s)&=&\frac{s^5}{1575(4\pi)^8}\left\{1+\frac{\alpha_s}\pi
  \left(2\ln\pfrac{\bar\mu^2}s+\frac{1021}{70}\right)\right\},\\
\rho_{\theta_2}(s)&=&\frac{s^5}{1575(4\pi)^8}\left\{1+\frac{\alpha_s}\pi
  \left(\frac58\ln\pfrac{\bar\mu^2}s+\frac{2963}{336}\right)\right\}
  \ =\ \rho_{\theta_3}(s).\quad\qquad
\end{eqnarray}
The perturbative correction to $\rho_{\theta_1}$ is the largest of all the
cases treated in this paper with $1021/70(\alpha_s/\pi)$. When the Lorentz
index is not contracted, we obtain an ordinary and a crossover contribution
for the correlators
\begin{equation}
\Pi^{\mu\nu}_{\theta_i}(x)=\langle 0|T\theta_i^{+,\mu}(x)
  \bar\theta_i^{+,\nu}(0)|0\rangle,\qquad i=2,3,
\end{equation}
which are the same for both currents, namely,
\begin{eqnarray}
\Pi^{\mu\nu,o}_{\theta_iB}(x)&=&-384x^2\slx(x^2g^{\mu\nu}-2x^\mu x^\nu)
  S_0(x^2)^5\left\{1+\frac{\alpha_s}\pi\left(\mu_x^2x^2\right)^\eps
  \left(\frac{10}{3\eps}+15\right)\right\}\nonumber\\&&\strut
  +256x^4\gamma^\mu\slx\gamma^\nu S_0(x^2)^5
  \frac{\alpha_s}\pi\left(\mu_x^2x^2\right)^\eps
  \left(\frac1\eps+\frac12\right),\\
\Pi^{\mu\nu,x}_{\theta_iB}(x)&=&384x^2\slx(x^2g^{\mu\nu}-2x^\mu x^\nu)
  S_0(x^2)^5\frac{\alpha_s}\pi\left(\mu_x^2x^2\right)^\eps
  \left(\frac1\eps-\frac12\right)\nonumber\\&&\strut
  -192x^4\gamma^\mu\slx\gamma^\nu S_0(x^2)^5
  \frac{\alpha_s}\pi\left(\mu_x^2x^2\right)^\eps
  \left(\frac1\eps+\frac12\right).
\end{eqnarray}
The total contribution is
\begin{eqnarray}\label{Pimunub}
\Pi^{\mu\nu}_{\theta_iB}(x)&=&-384x^2\slx(x^2g^{\mu\nu}-2x^\mu x^\nu)
  S_0(x^2)^5\left\{1+\frac{\alpha_s}\pi\left(\mu_x^2x^2\right)^\eps
  \left(\frac8{3\eps}+\frac{31}2\right)\right\}\nonumber\\&&\strut
  +64x^4\gamma^\mu\slx\gamma^\nu S_0(x^2)^5
  \frac{\alpha_s}\pi\left(\mu_x^2x^2\right)^\eps
  \left(\frac1\eps+\frac12\right).
\end{eqnarray}
The counterterm reads
\begin{equation}\label{Pimunud}
\Delta\Pi^{\mu\nu}_{\theta_i}(x)= \frac{\alpha_s}\pi\frac1\eps S_0(x^2)^5
  \left[896x^2\slx(x^2g^{\mu\nu}-2x^\mu x^\nu)
  -64x^4\gamma^\mu\slx\gamma^\nu\right].
\end{equation}
For the calculation of the absorptive part of~(\ref{Pimunub})
and~(\ref{Pimunud}) related to $x^2\slx(x^2g^{\mu\nu}-2x^\mu x^\nu)$ one has to
extend the considerations of Appendix~A to tensors of rank~3, resulting in a
spectral density
\begin{equation}
\rho_{\theta_i1}=\frac{s^4}{2520(4\pi)^8}\left\{1+\frac{\alpha_s}\pi
  \left(\frac73\ln\pfrac{\bar\mu^2}s+\frac{9613}{360}\right)\right\}.
\end{equation}
For the calculation of the absorptive part related to
$x^4\gamma^\mu\slx\gamma^\nu$ one can use Appendix~A directly to obtain
\begin{equation}
\rho_{\theta_i2}=\frac{s^5}{9450(4\pi)^8}\left\{\frac{\alpha_s}\pi
  \left(\ln\pfrac{\bar\mu^2}s+\frac{1129}{210}\right)\right\}.
\end{equation} 

\section{Conclusion}
The quark condensate $\langle\bar qq\rangle$ is the only dimensionful quantity
number that appears in the sum rule analysis since we assume factorization for
the vacuum expectation value of the six quark operators~\cite{Chetyrkin:yr}).
Inclusion of terms $\sim m_s$ should not substantially change the quantitative
results as the mass of the strange quark is small~\cite{strangemass}. It
therefore follows from dimensional arguments that the mass of the pentaquark
baryon is $m_\theta\sim(|\langle\bar qq\rangle|)^{1/3}$ as long as power
corrections determine the mass. Consequently, $m_\theta$ should not change by
more than $\sim 10\%$ if $\langle\bar qq\rangle$ varies by $30\%$. Such a
variation is quite possible because of the uncertainties in the light quark
masses determined from the numerical value of the light quark condensate as
calculated from the partially conserved axial current (PCAC) relation for the
pion. The analogous expression for the strange quark condensate obtains some
corrections due to the $s$-quark mass which are well under
control~\cite{Ovchinnikov:1985bs}. Nevertheless, this still leaves the
uncertainty whether the pentaquark state is above or below the threshold.

We recall in this respect that the high accuracy of the MIT quark-bag model
permitted Jaffe to conclude that the dibaryon state $H$ lies below the 
$\Lambda\Lambda$ threshold and is therefore stable. The same conclusion was
drawn from a model calculation based on chiral solitons in
Ref.~\cite{Diakonov:1997mm}. However, for a model independent approach, the
relatively low accuracy of the method in the determination of the mass
($\sim 15\%$) does not make it possible to draw any conclusion about whether
the mass of the exotic baryon lies below or above the $KN$ threshold.

In this paper we have calculated NLO perturbative corrections to the
correlator of various pentaquark currents and their absorptive parts. We have
shown that such a calculation can be done by purely algebraic means for any
given interpolating current using the modular methods developed by us in
detail. As it turns out, the NLO corrections to the correlators are large. As
the  coupling constant is large at the relevant energy scale~\cite{taudecay},
the large perturbative $\alpha_s$ corrections will heavily change the relative
weight of the perturbative and the nonperturbative condensate terms. It would
be interesting to find out how the large NLO corrections to the absorptive
parts of the current correlators affect the sum rule analysis of pentaquark
states. This would form the subject of a separate publication.

\subsection*{Acknowledgments}
This work was supported in part by the Estonian target financed project
No.~0180056s09, by the Estonian Science Foundation under grant No.~8769, by
the RFBR grant No.~11-01-00182-a and by the DFG grant under contract No.\
DFG~SI~349/10-1. S.~Groote and A.A.~Pivovarov acknowledge partial support from
the Deutsche Forschungsgemeinschaft (436EST~17/1/06 and 436RUS~17/68/06).

\begin{appendix}
\section{Explicit derivation of the spectral density}
\setcounter{equation}{0}\def\theequation{A\arabic{equation}}
The momentum space representation of the correlator function can be obtained
from the configuration space representation by using the integration formula
\begin{equation}
\Pi(p)=2\pi^{\lambda+1}\int_0^\infty\pfrac{px}2^{-\lambda}J_\lambda(px)
  \Pi(x)x^{2\lambda+1}dx
\end{equation}
where $\lambda=1-\eps$ and $J_\lambda(z)$ is the Bessel function of the first
kind. In the case that the correlation function $\Pi(x)=(x^2)^{-\alpha}$ is a
simple power in $x^2$, the integral can be obtained analytically. The result is
\begin{equation}
\Pi_\alpha(p^2)=\pi^{\lambda+1}\pfrac{p^2}4^{\alpha-\lambda-1}
  \frac{\Gamma(\lambda+1-\alpha)}{\Gamma(\alpha)}.
\end{equation}
The corresponding spectral density is given by the discontinuity of the
correlation function where the discontinuity of the correlation function (in
the Euclidean domain!) lies along the negative real axis. One obtains
\begin{equation}
\rho_\alpha(s)=\frac1{2\pi i}\Disc\Pi_\alpha(-s)=\pi^{\lambda+1}
  \pfrac{s}4^{\alpha-\lambda-1}\frac1{\Gamma(\alpha)\Gamma(\alpha-\lambda)}.
\end{equation}
In order to calculate the spectral density, we have to use the scalar
correlation function. The vector-type correlation function of the pentaquarks
(as well as those of all states composed of fermions) can be obtained from 
the derivative of this scalar correlation function $F(x^2)$. One has
\begin{equation}\label{deriv}
\partial_\mu F(x^2)=2x_\mu\frac{\partial F(x^2)}{\partial x^2}=x_\mu f(x^2).
\end{equation}
Given the function $f(x^2)$, the scalar correlation function is obtained by
integrating over $x^2/2$. In case of pentaquarks, we have
\begin{equation}
f(x^2)=\left(S_0(x^2)\right)^5(x^2)^2\left\{A+\frac{\alpha_s}\pi
  (\mu_x^2x^2)^\eps B\right\}
\end{equation}
where $A$ contains the LO contribution and the counterterm and where $B$
contains the NLO contribution. Recalling the $x^2$ dependence of $S_0(x^2)$ in
Eq.~(\ref{S0def}), one obtains
\begin{eqnarray}
F(x^2)&=&\frac12\pfrac{-\Gamma(2-\eps)}{2\pi^{2-\eps}}^5\int\left\{
  (x^2)^{5\eps-8}A+\frac{\alpha_s}\pi\left(\mu_x^2\right)^\eps(x^2)^{6\eps-8}B
  \right\}dx^2\ =\nonumber\\
  &=&\frac12\pfrac{-\Gamma(2-\eps)}{2\pi^{2-\eps}}^5\left\{
  \frac{(x^2)^{5\eps-7}}{5\eps-7}A+\frac{\alpha_s}\pi\left(\mu_x^2\right)^\eps
  \frac{(x^2)^{6\eps-7}}{6\eps-7}B\right\}.
\end{eqnarray}
For the corresponding spectral density one has
\begin{eqnarray}
\rho_F(s)&=&\frac12\pfrac{-\Gamma(2-\eps)}{2\pi^{2-\eps}}^5\pi^{2-\eps}
  \times\strut\nonumber\\&&
  \left\{\frac{(s/4)^{5-4\eps}A}{(5\eps-7)\Gamma(7-5\eps)\Gamma(6-4\eps)}
  +\frac{\alpha_s}\pi\left(\mu_x^2\right)^\eps\frac{(s/4)^{5-5\eps}
  B}{(6\eps-7)\Gamma(7-6\eps)\Gamma(6-5\eps)}\right\}\ =\nonumber\\
  &=&\frac{-\pi^{2-\eps}(s/4)^{5-4\eps}}{2\Gamma(8-5\eps)\Gamma(6-4\eps)}
  \pfrac{-\Gamma(2-\eps)}{2\pi^{2-\eps}}^5\times\strut\nonumber\\&&
  \left\{A+\frac{\alpha_s}\pi\pfrac{\bar\mu^2}s^\eps B
  \Big(1+\left(\psi(8)+\psi(6)+2\gamma_E\right)\eps\Big)\right\}\ =\nonumber\\
  &=&\frac{(s/4)^{5-4\eps}\Gamma(2-\eps)^5}{64\pi^{8-4\eps}
  \Gamma(8-5\eps)\Gamma(6-4\eps)}
  \left\{A+\frac{\alpha_s}\pi\pfrac{\bar\mu^2}s^\eps B
  \left(1+\frac{512}{105}\eps\right)\right\},\qquad
\end{eqnarray}
where we have made use of the expansion
$\Gamma(a-\eps)=\Gamma(a)\left(1-\eps\psi(a)+O(\eps^2)\right)$ and where we
have incorporated the scale change $\mu_x=e^{\gamma_E}\bar\mu/2$.
Here $\psi(a)=\Gamma'(a)/\Gamma(a)$ is the polygamma function. We then use
\begin{equation}
A=A_0+\frac{\alpha_s}\pi\left(\frac{C_0}\eps+C_1\right),\qquad
B=\frac{B_0}\eps+B_1
\end{equation}
where $A_0$ is the LO contribution, $B_0$ and $B_1$ are the singular
respectively finite NLO contribution, and $C_0$ and $C_1$ are the singular
respectively finite contribution of the counterterm ($C_0=-B_0$). One finally
obtains
\begin{eqnarray}
\rho_F(s)&=&\frac{(s/4)^{5-4\eps}\Gamma(2-\eps)^5}{64\pi^{8-4\eps}
  \Gamma(8-5\eps)\Gamma(6-4\eps)}\times\strut\nonumber\\&&
  \left\{A_0+\frac{\alpha_s}\pi\left(\frac{B_0+C_0}\eps
  +B_0\ln\pfrac{\bar\mu^2}s+\frac{512}{105}B_0+B_1+C_1\right)\right\}
  \ =\nonumber\\
  &=&\frac{s^5}{604800(4\pi)^8}\left\{A_0+\frac{\alpha_s}\pi
  \left(B_0\ln\pfrac{\bar\mu^2}s+\frac{512}{105}B_0+B_1+C_1\right)\right\}.
  \qquad
\end{eqnarray}
Because the singularities cancel one can set $\eps=0$ in the last step.

\section{On the choice for the interpolating current}
\setcounter{equation}{0}\def\theequation{B\arabic{equation}}
Formally, the QCD sum rule method dictates {\it a priori\/} the only condition
for the choice of the current: it must possess the required quantum numbers.
However, {\it a posteriori\/} (after the fit) positive results can be obtained
only for ``physical'' currents. In particular, the two requirements formulated
below are usually necessary. This may be an indication that there exists a
criterion which makes it possible to select the optimal (physical) current
from the set of currents with the quantum numbers of the given channel. This
argument was also given in Ref.~\cite{Larin:1985gb}, in which the deuteron
mass was calculated by the same method.

The choice of the interpolating current is crucial in this respect and has to
be considered very carefully. The physical current $\Theta$ satisfies the
following two requirements. First, there exists a nonzero nonrelativistic limit
for it (i.e.\ if the quark field $\psi(x)$ is represented in the standard
manner in terms of the small and large components, the term containing only
the large component will be nonzero). We note that in
Refs.~\cite{Ioffe,Ioffe:1981me} it was already pointed out that the existence
of a nonrelativistic limit is a desirable property for the construction of
currents when employing the QCD sum rule method. The demand for the existence
of such a limit is quite natural as the results should be reproducible in some
effective potential model of constituent quarks. Second, the colour (and
flavour) structure is important. This will be explained in some detail in the
following.

According to Ref.~\cite{Sasaki:2003gi}, ``the $qqqq\bar q$ state can be
decomposed into a pair of color singlet states as $qqq$ and $q\bar q$ [\dots].
For instance, one can start a study with a simple minded local operator for
the $\Theta^+(1540)$, which is constructed from the product of a neutron
operator and a $K^+$ operator such as
$\Theta=\epsilon_{abc}(d_a^TC\gamma_5u_b)d_c(\bar s_e\gamma_5 u_e)$. The
two-point correlation function composed of this operator, in general, couples
not only to the $\Theta$ state (single hadron) but also to the two-hadron
states such as an interacting $KN$ system. Even worse, when the mass of the
$qqqq\bar q$ state is higher than the threshold of the hadronic two-body
system, the two-point function should be dominated by the two-hadron states.
Thus, a specific operator with as little overlap with the hadronic two-body
states as possible is desired in order to identify the signal of the
pentaquark states [\dots].'' And following Ref.~\cite{Zhu:2003ba}, ``isospin
and color structure guarantee that these currents will never couple to a
$K^+n$ molecule or any other $K^+n$ intermediate state [\dots].'' This is the
reason to use a nontrivial colour structure in the previous paper.

We briefly comment also on a second choice of currents with nontrivial
flavour structure which was not considered in this paper. If we represent the
current $\Theta$ in the form of a product ``singlet$\otimes$singlet'' with
respect to colour, i.e., in the form $\Theta=\Psi(x)\Gamma\Phi(x)$, where each
of the operators $\Psi$, $\Phi$ is a colour singlet and $\Gamma$ is a string
of Dirac $\gamma$ matrices, we can choose
\begin{equation}
\Theta=\Psi_{A_3}^{A_4}(x)\Phi_{A_4}^{A_3}(x)
\end{equation}
where the colour-singlet operator $\Psi$ is a flavour octet,
\begin{equation}
\label{flavourOctet}
(\Psi_{A_3}^{A_4})^\alpha
  =\epsilon_{a_1a_2a_3}(\psi_{A_1}^{a_1} C\gamma_5 \psi_{A_2}^{a_2})
  (\psi_{B_1}^{a_3})^\alpha\frac12\epsilon^{A_1A_2B_2}\left(\delta_{A_3}^{B_1}
  \delta_{B_2}^{A_4}-\frac13\delta_{B_2}^{B_1}\delta_{A_3}^{A_4}\right)\,.
\end{equation}
Here $\alpha$ is a spinor index. The flavour octet~(\ref{flavourOctet}) has
the quantum numbers of the baryon octet and has e.g.\ been used in
Refs.~\cite{Ioffe,Chung:1981cc} to calculate the properties of light
baryons within the QCD sum rule method. Thus, if the current $\Theta$ is
represented in the form ``singlet$\otimes$singlet'' with respect to colour, it
has to have the structure ``octet$\otimes$octet'' with respect to flavour
(each colour singlet is a flavour octet). Physically, the following picture
emerges. The colourless state $\Theta$ splits into two colourless clusters
which separate at large distances to become a meson and a baryon. As a result,
we conclude that the current $\Theta$ is the most physical current in the
sense that it has a nonrelativistic limit and that it is constructed as a
``octet$\otimes$octet'' state with respect to flavour. 

One can introduce interpolating currents including space-time derivatives of
the field operators. This is, for instance, needed for the  description of 
the orbital excitations of hadrons. The calculational techniques developed in
this paper can also be applied to these cases. However, in this paper we 
have restricted our discussion to interpolating currents without derivatives.
 
\end{appendix}

\end{document}